\documentclass[twocolumn,showpacs,preprintnumbers,amsmath,amssymb,floatfix]{revtex4-1}

\usepackage{graphicx}
\usepackage{dcolumn}
\usepackage{bm}
\usepackage{epsfig}
\usepackage{epstopdf}
\usepackage{subfigure}

\begin{document}

\title{Correction-to-scaling exponent for two-dimensional percolation
}

\author{Robert M. Ziff}%
\email{rziff@umich.edu}
\affiliation{%
Center for the Study of Complex Systems and Department of Chemical Engineering, University of Michigan, Ann
Arbor MI 48109-2136.
}%


\date{Received: date / Accepted: date}

\begin{abstract}
We show that the correction-to-scaling exponents in two-dimensional percolation are bounded by $\Omega \le 72/91$, $\omega = D \Omega \le 3/2$, and 
$\Delta_1 = \nu \omega \le 2$, based upon Cardy's result for the critical crossing probability on an annulus.  The upper bounds are consistent with many previous measurements of site percolation on square and triangular lattices, and new measurements for bond percolation presented here, suggesting this result is exact.  A scaling form evidently applicable to site percolation is also found.  
\end{abstract}
\maketitle
In percolation, a quantity of central interest is the size distribution $n_s(p)$, which gives the number of clusters (per site)  containing $s$ sites, as a function of the site or bond probability $p$.  \  In the scaling limit, in which $s$ is large and $p - p_c$ is small such that $(p - p_c)s^\sigma$ is constant, $n_s(p)$ behaves as
\cite{StaufferAharony94}
\begin{equation}
n_s(p) \sim A s^{-\tau} f(B(p-p_c)s^\sigma) \ ,
\label{ns}
\end{equation}
where $\tau$, $\sigma$, and $f(z)$ are universal, while the metric constants $A$ and $B$  and the threshold $p_c$ are system-dependent.  For two-dimensional systems, $\tau = 187/91$, $\sigma = 36/91$. \  The constants $A$ and $B$ are uniquely defined if one assumes for example $f(0) = f'(0) = 1$ or $f(0)=1$ and $\int_{-\infty}^\infty f(z) dz = 1$.

Here we are concerned with finite-size corrections to Eq.\ (\ref{ns}) at $p_c$, where $n_s(p_c) \sim A s^{-\tau}$.  It is generally hypothesized that the leading  corrections to this behavior are of the form
\begin{equation}
n_s(p_c) \sim A s^{- \tau} (1 + Cs^{-\Omega} + \ldots)
\label{nsp}
\end{equation}
where $\Omega$ is the correction-to-scaling exponent.   This term corresponds to a correction in terms of a length scale $L \sim s^{1/D}$ of the form $(1 + c L^{-\omega} + \ldots)$, where $\omega = D \Omega$ and $D = 91/48$ is the fractal dimension.

\begin{table*}[htdp]
\caption{History of determinations of $\Omega$, $\omega = D \Omega = (91/48) \Omega$, and $\Delta_1 = \Omega/\sigma = (91/36) \Omega$. \ 
Numbers in parentheses
represent errors in last digit(s), and are shown on original values only.}
\begin{center}
\begin{tabular}{|l|l|l|l|l|l|l|l|}
\hline
year & author & method & $\Omega$ & $\omega$  & $\Delta_1$\cr 
\hline
1976 & Gaunt \& Sykes  \cite{GauntSykes76}  & series  & $0.75(5)$ & 1.42 & 	1.90  \cr
1978 & Houghton, Reeve \& Wallace \cite{HoughtonReeveWallace78} & field theory & 0.54--0.68 & 0.989--1.28 & 1.32--1.71\cr
1979 & Hoshen et al. \cite{HoshenStaufferBishopHarrisonQuinn79} & MC  & 0.67(10) & 1.27 & 1.69 \cr
1980 & Pearson \cite{Pearson80} & conjecture & $64/91 \approx 0.703$ & 1.333& 1.778 \cr
1980 & Nakanishi \& Stanley \cite{NakanishiStanley80}& MC & $0.6 \le \Omega \le 1$ & & \cr
1982& Nienhuis \cite{Nienhuis82}  & field theory & $96/91 \approx 1.055$ &2& 2.667 \cr
1982 & Adler, Moshe \& Privman  \cite{AdlerMoshePrivman82} & series& $0.5$ &0.95& 1.26\cr
1983 & Adler, Moshe \& Privman  \cite{AdlerMoshePrivman83} & series & $0.66(7)$ & 1.25& 1.67\cr
&\qquad '' & series $p < p_c$& $0.49$ & 0.93& 1.24\cr
 1983 & Aharony \& Fisher \cite{AharonyFisher83,MargolinaDjordjevicStaufferStanley83} & RG theory & $55/91 \approx 0.604$ & $55/48 \approx 1.15$ & $55/36 \approx 1.53$\cr
1984 & Margolina et al.\ \cite{MargolinaDjordjevicStaufferStanley83,MargolinaNakanishiStaufferStanley84} & MC & $0.64(8)$ & 1.21& 1.62\cr
&\qquad " & series  &  0.8(1) &1.52 & 2.02 \cr
 1985 &Adler \cite{Adler85} & series & 63(5) &1.19 &1.59 \cr
1986 & Rapaport\ \cite{Rapaport86} & MC & 0.71--0.74 & & \cr
1998 & MacLeod \& Jan \cite{MacLeodJan98} & MC  & 0.65(5) & 1.23 & 1.64 \cr
1999 & Ziff and Bablievski \cite{ZiffBabalievski99} & MC  & 0.77(2) & 1.46 & 1.95\cr
2001 & Tiggemann \cite{Tiggemann01} & MC  & 0.70(2) & 1.33 & 1.77 \cr
2007 & Tiggemann \cite{Tiggemann07} & MC  & 0.73(2) & 1.38 & 1.85 \cr
2008 & Kammerer, H\"ofling \& Franosch \cite{KammererHoflingFranosch08}&MC  & 0.77(4) & 1.46 & 1.95\cr
2010 & this work & theory &$72/91 \approx 0.791$ & 1.5 & 2 \cr
\hline
\end{tabular}
\end{center}
\label{table1}
\end{table*}%

Studies of $\Omega$ go back to the  mid-1970's and are summarized in Table \ref{table1}; this is an updated version of a table given 
in \cite{AdlerMoshePrivman83}, which surveyed the work up to 1983.  The values were found by Monte-Carlo
simulation, analyses of series expansions, and theoretical arguments.  
Note that Adler, Moshe and Privman  \cite{AdlerMoshePrivman82} found $\Omega = 0.66 \pm 0.07$ from an analysis of series at $p_c$, but found $\Omega \approx 0.49$
when studying the scaling behavior of the mean cluster size for $p < p_c$, where corrections are proportional to 
$(p_c - p)^{\Delta_1}$ and $\Delta_1$ is related to $\Omega$ by  $\Delta_1  = (\beta \gamma)\Omega =(91/36) \Omega $. \ Theoretical analyses of corrections to scaling were given by Aharony and Fisher \cite{AharonyFisher83} and Herrmann and Stauffer \cite{HerrmannStauffer84}.
Derrida and Stauffer  \cite{DerridaStauffer85} studied $\omega$ on strips, and their values for two different orientations extrapolate to a value of $\omega$ somewhere in the range of 1 to 2.
Bhatti et al.\ find $\Delta_1$ in the range $\approx 1 - 3$ for a variety of measurements related to backbones of 2D percolation clusters 
\cite{BhattiBrakEssamLookman97}.
In more recent work, Ziff and Bablievski \cite{ZiffBabalievski99} found  numerically that $\Omega = 0.77\pm 0.02$, and independently
Kammerer et al.\ \cite{KammererHoflingFranosch08} found the identical value, $\Omega = 0.77 \pm 0.04$.  A nearby value $0.73 \pm 0.02$
has also been given recently by Tiggemann \cite{Tiggemann07}.


In this Letter, we derive a value for $\Omega$ by relating Cardy's recent  result  \cite{Cardy06} for crossing in an annulus to the cluster size distribution.
For the probability $\Pi(\tau)$ that a crossing occurs in an annulus or a cylinder, Cardy found
\begin{equation}
\Pi(\tau) = \frac{\eta(-1/3 \tau) \eta(-4/3 \tau)}{\eta(-1/\tau)\eta(-2/3\tau)} =(3/2)^{1/2}\frac{\eta( 3 \tau ) \eta( 3 \tau / 4)}{\eta(\tau)\eta(3 \tau /2)}
\label{pitau}
\end{equation}
where $\eta(\tau) = q^{1/24} \prod_{n = 1}^\infty (1 - q^n) = \sum_{n=-\infty}^\infty (-1)^n q^{(6n + 1)^2/24}$  is the Dedekind $\eta$ function, with $q = e^{2 \pi {\mathrm i} \tau}$.  For an annulus of outer and inner radii $R$ and $R_1$ respectively,  $\tau = (\mathrm i/ \pi) \ln(R/R_1)$.  For a cylinder of circumference $\ell$ and length $L$, $\tau = 2 {\mathrm i} L/\ell $, since by conformal transformation $R/R_1$ corresponds to $e^{2 \pi L/\ell}$. 
Note that $\eta(\tau)$  can be directly evaluated in Mathematica using {\tt DedekindEta[$\tau$]}.
 
Eq.\  (\ref{pitau}) is consistent with several previous numerical measurements of crossing.   For a cylinder of aspect ratio $L/\ell = 1/2$, which corresponds to an annulus with $R/R_1 = e^{\pi}$, $\tau = \mathrm i$ and Eq.\ (\ref{pitau}) gives $\Pi( \mathrm i) \approx 0.876\,631\,451\ldots$, in agreement with the precise numerical value $0.876\,657(45)$ found by de Oliveira, N\'obrega, and Stauffer \cite{Oliveira03}. \  For $L/\ell = 1$ or $R/R_1 = e^{2 \pi}$, $\tau =  2 \mathrm i$ and Eq.\ (\ref{pitau}) gives $\Pi(2 {\mathrm i}) \approx 0.636\,454\,001$, which agrees closely with the measured values 0.636\,65(8) of Hovi and Aharony \cite{HoviAharony96}, 0.638 of Acharyya and Stauffer \cite{AcharyyaStauffer98}, 0.64(1) of Ford, Hunter, and Jan \cite{FordHunterJan99}, 0.6365(1) of Shchur  \cite{Shchur00}, and 0.6363(3) (average) by Pruessner and Moloney \cite{PruessnerMoloney03}.

Note that Eq.\ (\ref{pitau}) implies $\Pi = 1/2$ for $L/\ell \approx 1.368\,800$, and the maximum in $-\Pi'(L/\ell)$ [the inflection point of $\Pi(L/\ell)$] occurs at $L/\ell \approx 0.540\,652$, where $-\Pi'$ has value $0.522\,282$. In comparison, for a system with open boundaries, the maximum in $-\Pi'(L/\ell)$ occurs at $L/\ell \approx 0.523\,522$ with value $0.737\,322$  \cite{Ziff95,Ziff95a}, as follows from
Cardy's open-boundary crossing formula \cite{Cardy92}.

Expanding the $\eta$ functions in (\ref{pitau}) for large $L/\ell$, we find,
\begin{equation}
\Pi(\tilde q) = \sqrt{\frac{3}{2}} \,\tilde q^{5/48}\left( 1 - \tilde q^{3/2} + \tilde q^2 - \tilde q^{7/2} + 2 \tilde q^4 - \tilde q^{9/2}  - \ldots\right)
\end{equation}
where $\tilde q = R_1/R = e^{-2 \pi L/\ell}$ as in \cite{Cardy06}.  \  Thus, for the annulus, we have
\begin{eqnarray}
\Pi&(&R/R_1) = \sqrt{\frac{3}{2}} \,\left(\frac{R}{R_1}\right)^{-5/48} \Bigg\{1 - \left(\frac{R}{R_1}\right)^{-3/2}  \cr
&+& \left(\frac{R}{R_1}\right)^{-2} - \left(\frac{R}{R_1}\right)^{-7/2} + 2 \left(\frac{R}{R_1}\right)^{-4} - \ldots \Bigg\}
\label{eq:PiR}
\end{eqnarray}
and for the cylinder, we have 
\begin{eqnarray}
\Pi&(&L/\ell) = \sqrt{\frac{3}{2}} \,e^{-(5/24) \pi L/\ell}\Bigg\{ 1 - e^{-3 \pi L/\ell}  \cr
&+& e^{-4 \pi L/\ell} - e^{-7 \pi L/\ell}+ 2 e^{-8 \pi L/\ell}  - \ldots\Bigg\} 
\label{PiL}
\end{eqnarray}
Shchur \cite{Shchur00} has verified the leading term numerically, finding $0.654\,48(5)$ for the exponent $5 \pi /24 \approx 0.654\,498$, and the intercept of the asymptotic line in his Fig.\ 6 is consistent with the predicted coefficient $\sqrt{3/2} \approx 1.224\,745$.  Preussner and Moloney \cite{PruessnerMoloney03} also measure this coefficient and find 1.2217(4) for bond percolation and 1.2222(4) for site percolation on a square lattice, where the errors bars represent statistical errors of a single set of systems. 


For the annular result, we can imagine that the system is actually infinite with an inner boundary of radius $R_1$; then, $\Pi(R/R_1)$ is the probability that a cluster connected to that inner boundary has a maximum radius greater than or equal to $R$. \  That is, if, in the infinite system, the cluster  connected to the center extends beyond a circle of radius $R$, then in the annulus there will be a crossing cluster between the two circles of radii $R_1$ and $R$, and these two events will occur with the same probability. 
Thus $\Pi$ gives a measure of the size distribution of the clusters connected to the inner circle, where the size is characterized by the maximum cluster radius.  
Here, we relate this to the cluster size distribution by associating the inner radius to the discreteness of the lattice.  


Given a size distribution $n_s$, the probability that an occupied site is connected to a cluster of size greater than or equal to $s$ is given by:
\begin{equation}
P_{\ge s}(p) = \sum_{s' = s}^\infty s' n_{s'}  \approx \int_s^\infty s' n_{s'} ds' 
\label{Pgesp}
\end{equation}
Using (\ref{nsp}) for $n_s(p_c)$, we thus find that, at $p_c$,
\begin{equation}
P_{\ge s}(p_c)  \sim A' s^{2 - \tau} (1 + B' s^{-\Omega} + \ldots)
\label{Pgespc}
\end{equation}
where $2 - \tau = -\beta \sigma = -5/91$ in 2-d.
Because critical percolation clusters are fractal, $s$ are $R$ are related by
\begin{equation}
s \sim s_0(R/\epsilon)^D
\label{fractal}
\end{equation}
where $D$ is the fractal dimension, $s_0$ is a constant, and $\epsilon$ is of the order of the lattice spacing and represents the lower size cutoff of the system,
similar to a boundaryt extrapolation length \cite{Ziff96}.  Putting (\ref{fractal}) into (6), assuming $\epsilon = R_1$, we find that the probability a cluster has a radius greater than $R$ is given by
\begin{equation}
P_{\ge R}(p_c) \sim a R^{D(2 - \tau)} (1 + b R^{-\Omega D} + \ldots)
\label{PgeRpc}
\end{equation}
where $a$ and $b$ are constants.  By hyperscaling $D(2 - \tau) = D - d = -\beta /\nu = -5/48$.  \  Comparing this with (\ref{eq:PiR}), we see that  $\Omega D = \omega = - 3/2$, or 
\begin{equation}
\Omega = 3/(2D) = 72/91
\label{Omega}
\end{equation}
implying also $\Delta_1 =  \nu \omega = 2 $.

Alternately, if we put $(R/R_1) = (s/s_0)^{1/D}$ into (\ref{eq:PiR}), we find
\begin{eqnarray}
&P&_{\ge s}(p_c) = \sqrt{\frac{3}{2}} \,\left(\frac{s}{s_0}\right)^{-\frac{5}{91}} \Bigg\{1 - \left(\frac{s}{s_0}\right)^{-\frac{72}{91}}  \cr
&+& \left(\frac{s}{s_0}\right)^{-\frac{96}{91}}
- \left(\frac{s}{s_0}\right)^{-\frac{168}{91}}  + 2 \left(\frac{s}{s_0}\right)^{-\frac{192}{91}} -  \ldots \Bigg\}
\label{Pgess}
\end{eqnarray}
which gives the higher-order corrections also.  

In deriving (\ref{Pgess})  we have ignored finite-size corrections to (\ref{fractal}).   Say we have to next order
\begin{equation}
s \sim s_0(R/\epsilon)^D (1 + c R^{-x} + \ldots)
\end{equation}
then this will lead to a term of order $(s/s_0)^{-x/D}$ in the expansion of $P_{\ge s}(p_c)$, and where $x/D$ lies within the exponents 72/91, 96/91, $\ldots$ will determine its importance.  It is of course possible that $x/D < 72/91$, in which case that would be the dominant correction.  There have been studies in the past on the finite-size corrections to the radius of gyration \cite{MargolinaFamilyPrivman84}, but not to our knowledge to the maximum radius with respect to an arbitrary point within a cluster, which is needed here.
Another source of finite-size corrections are discussed by Fisher and Aharony (see footnote 18 of \cite{MargolinaDjordjevicStaufferStanley83}) which imply a correction $\Omega_0 = 55/91 = 0.604$.


Our result for $\Omega$ is consistent with the very first determination  $\Omega = 0.75 \pm 0.05$
in 1975 by Gaunt and Sykes, and many of the subsequent results including the latest numerical values.
Some previous results gave lower values, such 
as the series results of \cite{AdlerMoshePrivman83}, which were based upon studying
the scaling of quantities like the mean cluster size away from $p_c$.
Perhaps these lower values were due to analytic corrections which are relevant away from $p_c$.

In Eq.\ (\ref{Pgess}), we see that the next-order correction term is $s^{-\Omega_2}$ with $\Omega_2 = 96/91$.
This interestingly is exactly the value of $\Omega$
proposed by Nienhuis \cite{Nienhuis82}.  The closeness of this exponent to $\Omega$ would make
its determination numerically difficult, and in any case higher-order corrections in the relation  (\ref{fractal}) between $s$
and $R$ might mask this term.

\begin{figure}[htbp] 
   \centering
   \includegraphics[width=3.5in]{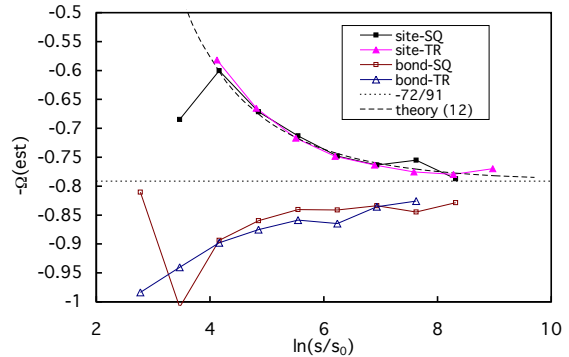} 
   \caption{Plot of  $-\Omega$(est) 
   as a function of $\ln (s/s_0)$ for $s = 2$, 4, $\ldots 1024$.  The two upper curves are a replotting of the data given in Fig.\ 6 of Ref.\ \cite{ZiffBabalievski99}, using $s_0 = 0.25$ (square lattice) and $s_0 = 0.13$ (triangular lattice) to line the data up with the
  theoretical prediction---the first three terms of  (\ref{Pgess}).  No vertical rescaling was done. (Note: there was a minus sign missing in the label of the vertical axis in Ref.\ \cite{ZiffBabalievski99}.)  The two lower curves are bond percolation on the square and triangular lattices, using $s_0 = 0.25$ and $0.5$, respectively. [Note: data will be updated for published version]}
   \label{fig}
\end{figure}

Assuming the other correction terms are small, Eq.\ (\ref{Pgess}) implies that
$P_{\ge s}$ should be a universal function of $s/s_0$, where $s_0$
varies from system to system.
This may explain an observation made in \cite{ZiffBabalievski99}
that the quantity $\Omega(\mathrm{est})$ defined by 
\begin{equation}
\Omega(\mathrm{est}) = -\log_2 \left(\frac{C_s - C_{s/2}}{C_{s/2}-C_{s/4}}\right)
\end{equation}
where $C_s = s^{\tau-2} P_{\ge s}$, when plotted vs.\ $\ln s$, appears to be a universal curve for 
site percolation on the square and triangular lattice, except for a horizontal shift.
$\Omega(\mathrm{est})$ equals $\Omega$
if the correction to scaling has only one term as in (\ref{Pgespc}), and otherwise can
be used to estimate $\Omega$ by taking $s$ relatively large.
  The shift that was needed in \cite{ZiffBabalievski99} to line up the data just reflects the
difference in $s_0$ between the two lattices, because of the logarithmic scale in the plot.
In Fig.\ \ref{fig} we have replotted
the data of Ref.\  \cite{ZiffBabalievski99} along with the results of using the $\Omega(\mathrm{est})$ that follow from (\ref{Pgess}) (using the first
three terms), and adjusted $s_0$ to match the theoretical prediction.  The behavior can be seen to match the theory very well,
with only a single adjustable parameter $s_0$. \  In Ref.\ \cite{ZiffBabalievski99}, we generated  $\approx 6 \cdot 10^9$ clusters up to size $s = 1024$ 
for each system, to obtain these data.

Here we have generated additional data for bond percolation on square and
triangular lattices, in which we characterize $s$ by the number of sites
wetted by the clusters.  (Evidently, these systems have not been studied
in the past to find $\Omega$.)   We generated $\approx 10^{11}$ clusters with $s \le 1024$ for each lattice,
using the R(9689) random-number generator of \cite{Ziff98}.
The corresponding $\Omega(\mathrm{est})$ are also plotted in Fig.\ \ref{fig}.  For both lattices 
$\Omega(\mathrm{est})$ approaches $\approx 0.8$ from above for larger
$s$, implying that (\ref{Omega}) is again the leading correction exponent, but for smaller $s$ the data do not follow the behavior implied by (\ref{Pgess}),
implying that for bond percolation, there are other significant finite-size corrections.  While (\ref{Pgess}) implies deviations from the leading behavior
are negative, for bond percolation in fact those deviations are positive.   Similar positive finite-size effects
are seen in the closely related SIR epidemic model  \cite{TomeZiff10}.

It appears that $\Omega$(est) gives a lower bound to $\Omega$ with site percolation, and an upper bound with 
bond percolation, both bracketing 0.79.  Evidently larger clusters are needed for $\Omega$(est) to 
get close to that value, but statistical errors limit the cluster size that can practically be used.
(The difficulty in reaching this asymptotic regime for the corrections might explain the
generally low values of $\Omega$ found in the past.)
We have also carried out an analysis of the exact $n_s(p_c)$ for site percolation on 
the square lattice, based upon the known series (\cite{SykesGlen76,Mertens90}) for $n_s(p)$ for $s \le 22$.
Results are lead to similar conclusions and will be discussed elsewhere.

We can find the behavior of $n_s$ from (\ref{Pgess}) as follows:
\begin{eqnarray}
&n_s&(p_c) = \frac{p_c}{s}\left(P_{\ge s} - P_{\ge (s+1)}\right)=  \frac{5 p_c}{91 s_0^2
} \sqrt{\frac{3}{2}} \left(\frac{s}{s_0}\right)^{-\frac{187}{91}}  \cr &\times& \Bigg\{1 - \frac{77}{5} \left(\frac{s}{s_0}\right)^{-\frac{72}{91}}-\frac{48}{91} \left(\frac{s}{s_0}\right)^{-1}+\frac{101 }{5} \left(\frac{s}{s_0}\right)^{-\frac{96}{91}}+\ldots\Bigg\} \cr
 \end{eqnarray}
 where the factor of $p_c$ is included for site percolation only \cite{StaufferAharony94}.   Thus, for Eq.\ (\ref{ns}), we have $A = (5 p_c/91)\sqrt{3/2} \, s_0^{5/91}$ 
 and $B = (77/5) s_0^{72/91}$, and here we have picked up the analytic term $s^{-1}$.  Likewise,
\begin{equation}
n_{\ge s}(p_c) =   \frac{5 p_c}{96 s_0} \sqrt{\frac{3}{2}} \left(\frac{s}{s_0}\right)^{-\frac{96}{91}}  \Bigg\{1 - \frac{44}{5} \left(\frac{s}{s_0}\right)^{-\frac{72}{91}}+\ldots\Bigg\} 
 \end{equation}
 For example for site percolation on the triangular lattice ($p_c = 0.5$), we found that $s_0 = 0.13$ (see Fig.\ \ref{fig}), which implies a coefficient of $0.0285$ to $n_{\ge s}(p_c)$.  This value agrees closely with the value $\approx 0.028$ found in \cite{MargolinaDjordjevicStaufferStanley83} (their Fig.\ 2.)  
 
 In conclusion, we found that the behavior of crossing on an annulus implies the correction-to-scaling exponent (\ref{Omega}) to the size distribution $n_s$, which seems to represent the dominant term for the systems we have studied, in comparison to other possible finite-size terms that might appear.   For future work, it would be interesting to study $n_s(p_c)$ on different lattices, and to explore directly the relation between $R$ and $s$ andfind higher-order corrections to (\ref{fractal}).

The author thanks Amnon Aharony and Joan Adler
for providing comments and references, and especially thanks Dietrich
Stauffer for many helpful suggestions.



\bibliographystyle{apsrev4-1}

\bibliography{ZiffPercFinite11arXiv}

%
%

\end{document}